\documentclass{article}
\usepackage{fullpage}
\newcommand{\comment}[1]{}
\newcommand{\half}[0]{\frac{1}{2}}

\newcommand{\ket}[1]{|#1\rangle}
\newcommand{\cket}[1]{|#1)}

\newcommand{\ketbra}[2]{|#1\rangle\langle #2|}
\newcommand{\cketbra}[2]{|#1)(#2|}

\begin{document}
\title{On Classical Teleportation  
and Classical Nonlocality}
\author{Tal Mor \\
\small  Computer Science Department, Technion,
Haifa 32000, Israel.}
\maketitle
\begin{abstract}

An interesting protocol for classical teleportation of an unknown
classical state was recently suggested by Cohen, and by Gour and Meyer.  
In that protocol, Bob can sample from a probability distribution 
${\cal P}$ that is given to Alice, even if Alice has absolutely no
knowledge about ${\cal P}$.
Pursuing a similar line of thought, 
we suggest here a limited form of nonlocality ---
``classical nonlocality''. Our nonlocality is 
the (somewhat limited) classical analogue
of the Hughston-Jozsa-Wootters (HJW) quantum nonlocality.
The HJW nonlocality (also known as ``quantum remote steering'')
tells us how, for a given density
matrix $\rho$, Alice can generate any $\rho$-ensemble on the
North Star. This is done using surprisingly few resources --- one 
shared entangled state (prepared in advance), 
one generalized quantum measurement, and {\em no} communication.
Similarly, our classical non locality (which we call
``classical remote steering'') presents how, for a given
probability distribution ${\cal P}$, Alice can generate
any ${\cal P}$-ensemble on the North Star, using only one correlated
state (prepared in advance), one (generalized) classical measurement, 
and {\em no} communication.

It is important to clarify that while the classical teleportation 
and the classical non-locality protocols are probably rather 
insignificant from a 
classical information processing point of view, 
they significantly contribute 
to our understanding of what exactly is quantum in their well 
established and highly famous quantum analogues.

\end{abstract}

\section{Introduction and Notations}
\label{sec:intro}

Processing information using quantum two-level systems (qubits),
instead of
classical bits, has led to many surprising results, 
such as quantum algorithms that are exponentially faster 
than the best known classical algorithm (e.g., for factoring large numbers),
teleportation of unknown states~\cite{BBCJPW93},
and quantum cryptography.
Another cornerstone of quantum foundations and quantum
information processing is the Einstein Podolski Rosen (EPR) 
paradox~\cite{EPR}, or better stated --- the EPR nonlocality, which was
later on generalized~\cite{Gisin89,HJW93} by Gisin
and by Hughston, Jozsa and Wootters (HJW).
Are all of these effects purely quantum, or do we need to look
more carefully into the details of each effect to see where the
``quantumness'' plays a role?  

A quantum state $\rho$ (whether pure or mixed), once measured,
yields a classical 
probability distribution over some possible classical
outcomes: 
${\cal P} = \{p_i, i\}$, namely, the result $i$ is obtained
with probability $p_i$, and $\sum_i p_i =1$. 
If measured in another basis, the same quantum state yields 
another probability distribution  
${\cal P}' = \{p_{i'},i'\}$ (with $\sum_{i'} p_{i'} =1$). 
A measurement of a quantum state in a particular basis is like
sampling {\em once} from the 
appropriate classical probability distribution.

Here we compare quantum states to classical probability
distributions, in order to improve our understanding 
of ``quantumness'' versus ``classicality''.
We show that classical probability distributions 
present some interesting phenomena that are closely
related to teleportation and nonlocality.

\paragraph{``Classical states'' notations:}

Let us first recall that for any (nonpure) density matrix $\rho$,  
one can write $\rho = \sum_i p_i \rho_i$ in
infinitely many ways. Each of these ways is described by the ensemble
of states $\{p_i,\rho_i\}$, 
and is conventionally called a
``$\rho$-ensemble'' of the density matrix $\rho$.
Even if the states $\rho_i$ are all pure states
($\rho_i = \ketbra{\psi_i}{\psi_i}$), 
there are still infinitely many different
ensembles of the type 
$\{p_i,\rho_i\}$ that describe the same density matrix $\rho$.
For instance, $\half(\ketbra{0}{0}+\ketbra{1}{1}) = \half(\ketbra{+}{+}+
\ketbra{-}{-})$, with $\ket{\pm}=(\ket{0}\pm\ket{1})/\sqrt{2}$.

We denote a classical state of a bit, ``0'' and ``1'', by the notation
$\cket{0}$ and $\cket{1}$, and when we mix such states we use the notation
$\cketbra{0}{0}$ and $\cketbra{1}{1}$, for consistency with quantum states
notations.
Let us refer to an arbitrary classical probability distribution by the
name ``classical state''. 
For instance, the classical (mixed) state of
an honest coin is ${\cal P}_{\rm honest-coin} =
\half \cketbra{0}{0} + \half \cketbra{1}{1}$.
The state of an arbitrary coin is 
${\cal P}_{\rm coin} =
p \cketbra{0}{0} + (1-p) \cketbra{1}{1}$.
Similarly, the state of an honest die is
${\cal P}_{\rm honest-die} = \sum_{i=1}^6
(1/6) \cketbra{i}{i}$,
and the state of an arbitrary die is 
${\cal P}_{\rm die} = \sum_{i=1}^6
p_i \cketbra{i}{i}$  with $\sum_i p_i = 1$. 
In this classical case --- a measurement (e.g., a single sampling) 
yields one of the outcomes
$\cket{i}$ with the appropriate probability $p_i$.
This is similar to the case of a special quantum state which is diagonal 
in the computation basis, and is measured in that basis.
We use the term ``generalized coin'' to specify a coin, or a die, or any other
system whose classical state is a given probability distribution.

In the above, we wrote each classical state as a mixture
of pure classical states $\cket{i}$. This presentation is
unique, as there is no choice of basis as in the quantum case.
Note that, unlike an unknown quantum state that cannot be cloned,
a classical state is defined in only one basis, 
and therefore can be cloned if {\em sampling with replacement} 
is allowed.

This presentation of the classical state (via the $\cket{i}$ basis) 
does not describe the most general way of mixing classical states.
For instance, the totally-mixed state describing an
honest coin can be made by mixing
(via an equal mixture)
two dishonest coins, one with probability $p$ of being 
$\cket{0}$ and the other with probability 
$1-p$ of being $\cket{0}$.
In general, any classical state ${\cal P}$ 
can be written as being made of an ensemble of
classical probability distributions (namely a mixture of
classical states)
${\cal P} = \sum_j p_j {\cal P}_j$. 
As in the quantum analogue, one can write 
${\cal P}$ in infinitely many ways, 
and each of these ways is described by an ensemble
of classical states $\{p_j,{\cal P}_j\}$. 
Following the quantum case, 
we call each of these ways a
``${\cal P}$-ensemble'' of the classical probability distribution
(namely, the classical state) ${\cal P}$.


\paragraph{The structure of this paper is as follows:}

In Section~\ref{sec:C-teleport} we present a recent 
result regarding teleportation: 
In the original teleportation scheme, 
an unknown quantum state can be teleported, via a shared 
maximally-entangled state,
and two bits of classical communication~\cite{BBCJPW93}. 
In a more limited sense, an 
unknown classical state  
can also be teleported, and this is done via a shared classical state, 
and one bit~\cite{Cohen03,GM05}.

In Section~\ref{sec:Q-nonlocal} we 
present the EPR~\cite{EPR} and
the HJW~\cite{HJW93} nonlocalities. In brief,
let Alice (who is in Haifa) and Bob (who is far away, say on the
North Star)
be two parties.
For any density matrix $\rho$, 
one can write $\rho = \sum_i p_i \rho_i$ in
many ways using 
various $\rho$-ensembles.
The HJW nonlocality tells us the following:
If Alice and Bob  
share a pure entangled state such that the
reduced density matrix in Bob's hands is 
$\rho_{{}_{\rm Bob}}$, 
then Alice
can generate for Bob (nonlocally) 
any ensemble of quantum states $\{p_i; \rho_i\}$
(where $p_i$ is the probability of $\rho_i$) 
as long as 
$ \sum_i p_i \rho_i =  
\rho_{{}_{\rm Bob}}$.
This generation of any desired $\rho$-ensemble
is done without any communication between Alice and Bob;
all Alice needs to do is to perform an appropriate 
(generalized) measurement
on her part of the shared entangled state.
The EPR nonlocality 
can be viewed as a
special case in which the shared state is a singlet 
and Alice chooses a standard measurement in the $z$ basis or the
$x$ basis.

Then, we present in Section~\ref{sec:C-nonlocal}
a new type of nonlocality --- a ``classical nonlocality''. 
As was already mentioned, any classical state,
${\cal P} = \sum_i q_i \cketbra{i}{i}$, 
can also be written as a mixture of classical states 
${\cal P} = \sum_j p_j {\cal P}_j$, in 
many ways using 
various ${\cal P}$-ensembles,
$\{p_j,{\cal P}_j\}$.   
Our ``classical nonlocality'' argument is as follows:
If Alice and Bob  
share a correlated classical state such that the
resulting state in Bob's hands 
(namely, Bob's marginal probability distribution) is 
${\cal P}_{{}_{\rm Bob}}$, 
then Alice
can generate for Bob (nonlocally) 
any ensemble of classical states $\{ p_j; {\cal P}_j\}$
(where $p_j$ is the probability of ${\cal P}_j$) 
as long as 
$ \sum_j p_j {\cal P}_j =  
{\cal P}_{{}_{\rm Bob}}$.
This generation of any desired ${\cal P}$-ensemble
is done without any communication between Alice and Bob;
all Alice needs to do is to perform an appropriate 
(generalized) measurement
on her part of the shared correlated state.

Note the trivial special case in which 
Alice measures the shared state in the pure-state basis, just by 
looking at her part. Then, with probability $p_i$,
she obtains the result $i$ and she knows with certainty
that Bob's resulting pure state is
$\cket{i}$. As this presentation (via pure classical states)
is unique, there is no classical analogue
to the EPR nonlocality.

\section{Classical teleportation of classical states}
\label{sec:C-teleport}

Consider a classical coin in an unknown classical state, namely
an unknown probability distribution ${\cal P}_{\rm coin} = 
p \cketbra{0}{0} + (1-p) \cketbra{1}{1}$. 
The probability of a ``head'' is 
$p$. Charley gives this coin to Alice (on earth) and she would like to
teleport it to Bob.
Here we mean that Bob will be able to 
flip the coin once and obtain a result according to the correct
probability distribution.  
Namely, Bob can sample the probability distribution {\em once}.

We would like to avoid the trivial solution in which Alice 
samples the probability distribution {\em once} 
and tells Bob the outcome. 
Note that, in this solution, by hearing Alice's bit Bob is forced 
to sample the probability distribution,
which is somewhat different from what we are looking for.
Instead, we would like to obtain a situation in which 
Bob holds the state 
${\cal P}_{\rm coin} = p \cketbra{0}{0} + (1-p) \cketbra{1}{1}$, 
and is able to sample it whenever he wants. 
Furthermore, note that if Alice samples the state and tells
Bob the outcome, then Alice, and also any other
receiver of that data, 
will be in the same position as Bob. Namely, they will all share Bob's data.
Charley  would like to see 
a scenario in which only Bob can sample  
the coin he (Charley) gave to Alice, and that Bob can do so 
(once only) whenever he wishes to.
[This situation then resembles quantum teleportation after which only Bob
holds the state and can measure it whenever he wishes to.]

What are the minimal resources required for that operation?
Will it help Alice and Bob if some data, unrelated to $p$, 
was shared in advance?

As found by~\cite{Cohen03,GM05}, 
Alice can indeed classically teleport the state of the coin to Bob
with infinite accuracy by sharing one bit with Bob in advance, 
and sending one bit:
Let us assume that Alice and Bob share one classical bit, which
is $0$ with probability half, and $1$ with probability half,
but they {\em do not know} its value. 
Namely, they share the state 
$${\cal P}_{\rm correlated}= \sum_{i=0}^1 \half \cketbra{i}{i}_{\rm Alice}
\otimes \cketbra{i}{i}_{\rm Bob} \ . $$
Let Alice flip the coin she got from Charley.
Alice then measures the eXclusive OR (XOR) of the values of 
the coin given by Charley, and her part of the correlated state 
${\cal P}_{\rm correlated}$
(but without learning each one separately). For instance, she 
could give Charley her part of the correlated state; Charley then would
flip his coin (instead of Alice)
to obtain one bit, look at Alice's secret bit, 
and tell her just one bit of information ---
the parity of the two bits.    
This parity is then sent to Bob. If it is ZERO, Bob 
does nothing; if it is ONE
he flips his part of the shared correlated state.
In both cases, he now holds the state 
${\cal P}_{\rm coin}$ 
of Alice's coin.
The probability of viewing $0$ or $1$ is identical to 
the probability determined by the coin Charley gave to Alice.

It is important to mention that the main 
difference between this and the case
where Alice flips the coin and tells Bob the result is 
that here only Bob (in addition to Charley, of course) has
the correct distribution, and no other person will. Not even Alice.
In addition, here, 
even Bob did not yet (at the end of the protocol) sample 
the coin. 
Such a protocol is interesting since it emphasizes the quantumness of the
quantum teleportation by showing that not all its aspects are quantum.
In particular, it helps in demystifying quantum teleportation, showing
that the ability to teleport an {\em unknown} state has (in some sense) a 
classical analogue.
This protocol also emphasizes 
the differences between {\em having} a state and {\em sampling} 
(measuring) a
state, in both the classical and quantum domains.  

Is this classical teleportation protocol interesting from
a classical information processing point of
view?  Probably not.  The interplay between Alice, Bob, 
and Charley as described above
is very similar to a trivial classical analogue --- 
the ``one-time pad''.
In the one-time pad protocol Alice and Bob share one secret bit,
and when Alice wishes to send Bob a bit (potentially, 
a secret bit given to her by Charley), 
she sends Bob the parity of that new bit
and the shared bit.
This way, only Bob can learn Alice's bit so it is a secret to others.
This conventional description of the one-time pad ignores the
possibility (which is at the heart of the classical teleportation protocol)
that Alice's new bit is not fixed by Charley, but instead, it
is sampled at random from some fixed probability distribution.

It is important to note that Cohen~\cite{Cohen03} 
uses very different
notations, and also defines the classical states in a way that makes them
unclonable, due to sampling a probability distribution 
{\em without replacement}.

\section{Quantum Nonlocality (Quantum Remote Steering)}\label{sec:Q-nonlocal}

We first describe the EPR (Bohm's version) nonlocality.
Alice and Bob share a fully entangled state of two qubits,
say $[\ket{00}+\ket{11}]/\sqrt{2}$, 
and Alice measures her qubit in one basis ($z$) or a conjugate
one ($x$).  Whatever result she obtains, she now knows Bob's
state.  This is true even if Alice and Bob are space-like
separated, so that no information can go from one to the other.
The ``paradox'' is obtained if one assumes that quantum states
are physical realities that cannot change faster than light,
since then it seems that Bob's state must be well defined in
both the $z$ basis and the $x$ basis, contradicting the
uncertainty principle. Once one is willing to accept that
quantum states are not local physical realities, the
paradox is changed into a nonlocality argument:
In the case in which Alice measures in the $z$
basis she determines from far away that Bob's state
(which is the completely mixed state, as far as Bob can 
test on his own) will be made of the
states $\ket{0}$ and $\ket{1}$, and she then also 
knows which of the two states it is. 
Alternatively, in the case in which Alice measures in the $x$ basis  
she determines from afar that Bob's state
will be composed of $\ket{+}$ and $\ket{-}$.
Thus, Alice can determine the ensemble of states from which Bob's density matrix
will be constructed, and furthermore, 
Alice can tell which specific state will appear at a particular
instance of running the protocol. 
Obviously, Alice could have made any standard~\footnote{ 
These measurements are known as complete measurements, 
or von-Neumann measurements.} 
projection measurement as well
on her qubit,
and thus enforce the appropriate distribution on Bob's state.
This type of nonlocality does not contradict Einstein's
causality principle, in the sense that  
information is not transmitted {\em faster than light}.

The HJW generalization is done by having  Alice and Bob share any 
pure entangled state, and by letting Alice perform any 
generalized measurement (POVM)~\cite{Peres88,Peres93}. 
Generalized measurements
are equivalent to standard projection measurements when 
the latter are performed 
on an enlarged system containing the original system plus an
ancilla in a known initial state~\cite{Peres93}.
Now, the HJW nonlocality argument is the following:
Let Alice and Bob  
share a pure entangled state $\ket{\psi_{\rm AB}}$, such that the
reduced density matrix in Bob's hands is $\rho_{\rm Bob}$, 
namely, 
$$\rho_{\rm Bob} = \rm{Tr}_{\rm Alice}
[\ketbra{\psi_{\rm AB}}{\psi_{\rm  AB}}]\ .$$
Alice
can generate for Bob (nonlocally) 
any ensemble of quantum states $\{p_i; \rho_i\}$
(where $p_i$ is the probability of $\rho_i$) 
such that 
$ \sum_i p_i \rho_i = \rho_{\rm Bob}$. 
Alice can generate
this ensemble in Bob's hands 
in the sense that she chooses which ensemble to generate, 
and after she prepares her desired  
$\rho$-ensemble, she can tell us (or tell Bob) which of the 
states ($\rho_i$) in the ensemble he has in his hand. The preparation 
of the desired $\rho$-ensemble is done by choosing the
appropriate generalized measurement and performing it in her
lab (on earth).

Quantum teleportation can be thought of as a special case
of HJW nonlocality
in which the shared state is a singlet and the 
ensemble is built from the teleported 
qubit-state $\alpha \choose \beta$ and its
rotations around the three axes~\cite{telePOVM1,telePOVM2}.
In that case Alice also tells Bob which rotation to perform 
in order to get the state $\alpha \choose \beta$, 
while both of
them do not need to know $\alpha$ and $\beta$ (e.g.,
if the qubit is provided to Alice by Charley).

\section{Classical Nonlocality (Classical 
Remote Steering)}\label{sec:C-nonlocal}

In order to search for the classical analogue of the above nonlocality,
let us note that, for pure entangled states, we are always promised that a basis
exists such that the state can be written as 
$\ket{\psi_{\rm AB}} = \sum_i \alpha_i \ket{i_{\mbox{Alice}}}
\ket{i_{\mbox{Bob}}} $ (with $\alpha_i$ the normalized amplitudes) 
via the Schmidt decomposition (see, for instance~\cite{Peres93}).
If this state decoheres so that nondiagonal terms become zero, the density matrix 
$\ketbra{\psi_{\rm AB}}{\psi_{\rm AB}} $
changes to 
$ \sum_i P_i \ketbra{i_{\mbox{Alice}}}{i_{\mbox{Alice}}} \otimes
\ketbra{i_{\mbox{Bob}}}{i_{\mbox{Bob}}} $ 
with 
$ p_i =  |\alpha_i|^2$.

The classical analogue of the HJW quantum nonlocality
is based on Alice and Bob sharing the classical analogue of
the decohered state
$$\rho_{\rm AB} = 
\sum_i p_i \ketbra{i}{i}_{\rm Alice} \otimes
\ketbra{i}{i}_{\rm Bob} \ . $$ 
The classical state can then be called ``fully correlated'' 
because each party can measure
his part of the shared state and learn with probability 1 the state of the other
party.
The state is written as 
$${\cal P}_{\rm fully\mbox{-}correl}= \sum p_i \cketbra{i}{i}_{\rm Alice}
\otimes \cketbra{i}{i}_{\rm Bob} \ .$$ 
Bob's ``reduced state'' (Bob's marginal probabilities) is then given by 
$$ {\cal P}_{\rm Bob} = \rm{Tr}_{\rm Alice}
[{\cal P}_{\rm fully\mbox{-}correl}] = \sum p_i \cketbra{i}{i}_{\rm Bob} \ .
$$

In order to study what Alice can do with the shared state, 
we start the analysis by considering the EPR nonlocality. This nonlocality
has no nontrivial classical analogue.
The reason is that in a quantum world, there is more than one basis,
while classically there is only one. Thus, if Alice is
only allowed to perform measurements on the system (without an
ancilla), the quantum case is very interesting as the possibility
for various measurements leads to the EPR nonlocality,
while the classical case is trivial:
Alice and Bob will share the fully correlated state,
${\cal P}_{\rm fully\mbox{-}correl},$ 
with $p_i$ equal half (for $i=0$ and $i=1$).
Alice will measure in the classical basis, which is the only allowed
basis, and thus will know $i$.

A slightly more complicated scenario is obtained if we let Alice choose
between doing nothing with probability $q$, or measuring her part
of the shared state, with probability $1-q$.
Then, the following  
ensemble is created,
${\cal P}_{{}_{\rm B}} = q {\cal P}_{\rm totally\mbox{-}mixed} + [(1-q)/2] \cketbra{0}{0}
+[(1-q)/2] \cketbra{1}{1}$,
with 
${\cal P}_{\rm totally\mbox{-}mixed} = 1/2 \cketbra{0}{0}
+1/2 \cketbra{1}{1}$. Note that Alice will know which of the three 
possible states is now held by Bob.
Note also that the choice that 
Alice made of whether to measure or do nothing 
can be mimicked by
flipping an appropriate dishonest coin  that has 
a probability $q$ for a ``head''.
Note also that Alice could actually view all cases, 
and with probability $q$
take an obtained result and ``forget'' it. 
This way she also obtains three sets
as before.

Once we allow Alice to add another system, even though it is a 
classical one, the situation becomes less trivial, and we can make
a similar connection between nonlocality and teleportation as was made 
in~\cite{telePOVM1,telePOVM2} for the quantum case. 
Let Alice and Bob share the state 
${\cal P}_{\mbox{fully-correl}}$.
What can Alice do if she wants Bob to
hold a dishonest coin with probability $p \ne \half$ of 
yielding one result and probability $1-p$ of yielding the
other? Namely, she wants him to sample (once) 
from a probability distribution  
${\cal P}_{0} = p \cketbra{0}{0} + (1-p) \cketbra{1}{1}$,
or from a probability distribution  
${\cal P}_{1} = (1-p) \cketbra{0}{0} + p \cketbra{1}{1}$, 
and to know which probability distribution he sampled from.
Alice will take such a coin (with probability distribution 
${\cal P}_{0}$), flip it, and observe the parity
of this coin and her part of the shared state 
{\em without
looking at each bit separately}.
Bob's probability distribution, conditioned on Alice observing $0$, is
${\cal P}_0 =
p \cketbra{0}{0} + (1-p) \cketbra{1}{1}$. On the other hand, 
Bob's probability distribution, conditioned on Alice observing $1$, is
${\cal P}_1 =
p \cketbra{1}{1} + (1-p) \cketbra{0}{0}$.
As in the previous example, also here Alice could actually measure both her 
crooked coin and her part of the shared bit, but she must forget the results
and remember just the parity when she splits the outcomes into two sets.

Her success in predicting Bob's resulting state can, of course, 
be verified: If Charley gave Alice the coin, and only he
knows $p$, and they repeat the experiment many times, it is easy for Alice to
convince Charley that she controls Bob's probability distribution, that
it is sometimes 
${\cal P}_0$ and sometimes 
${\cal P}_1$, and that she knows when is which.
Note that this example of a ``classical nonlocality
protocol'' requires no communication between Alice and Bob.

Let us look at the case in which communication is added. 
Alice tells Bob the parity bit,
Bob rotates his part of the shared state if the parity
bit he learns equals 1 (or does nothing if that parity equals 0),  
and they then recover the classical teleportation protocol
of Section~\ref{sec:C-teleport}. Note that Alice and Bob could 
apply the protocol even if Alice does not know the classical state.

In the more general case, Alice can use a larger space (e.g., flip several 
coins together with the shared bit, or throw a die together with her part of 
the shared state) 
and predict one out of many (instead of one out of two) probability
distributions.

In order to deal with the most general case of this classical nonlocality, 
let us rephrase the ``classical teleportation'' 
case we have just seen (without the communication step).  
Let Alice replace the calculation of the parity by a different method:
she looks at the bit she shares with Bob, if it is $\cket{0}$ she flips the coin,
and if is it $\cket{1}$ she switches the numbers ``0'' and ``1'' on the coin, 
and flips the coin.
Now she looks at the result of the coin and ``forgets'' whether she had seen 
$\cket{0}$ or $\cket{1}$ on the shared bit.
Clearly, if the result of the flipped coin is ``0'', Bob's probability
distribution is ${\cal P}_0$ and if the result  
of the flipped coin is ``1'', Bob's probability
distribution is ${\cal P}_1$. 
By generalizing this method we will now obtain the proof of the most general case of
our classical nonlocality.

\paragraph{A protocol presenting classical nonlocality:}

Let Alice and Bob share the fully correlated state
${\cal P}_{\rm fully\mbox{-}correl}= \sum p_i \cketbra{i}{i}_{\rm Alice}
\otimes \cketbra{i}{i}_{\rm Bob}$.
Then Bob's reduced state (Bob's marginal probability
distribution) is  
${\cal P}_{{}_{\rm Bob}} 
= \sum p_i \cketbra{i}{i}_{\rm Bob} $ 
when written as the (unique) mixture of pure classical states $\cket{i}$.
Alice can generate any 
${\cal P}$-ensemble on the
North Star, provided that the ensemble satisfies
$ \sum_j p_j {\cal P}_j = {\cal P}_{{}_{\rm Bob}} $.
Let $I$ be the set of events $i$: the fully correlated state is sampled and it
provides the value $i$,
with probability $p_i$.
Let $J$ be the set of events $j$: with probability $p_j$
Bob has the classical state ${\cal P}_j$.

Let us write each of Bob's states  
$ {\cal P}_j $ via the basis states as
$ {\cal P}_j = \sum_i p_{(i|j)} \cketbra{i}{i}_{\rm Bob}  $ 
where, for each $j$, $ \sum_i p_{(i|j)} =1$, and 
$\sum_j p_j p_{(i|j)} = p_i$.
To create her desired 
${\cal P}$-ensemble, Alice follows this protocol:
\begin{enumerate}
\item
Alice looks at the shared state to find $\cket{i}$.
\item
For each result $\cket{i}$ Alice throws a generalized coin\footnote{
A coin with $j$ outcomes.} 
distributed according to 
$ p_{(j|i)} $, 
where these probabilities are obtained from the known 
probability distributions $p_i$, $p_j$, and $p_{(i|j)}$ using the Bayes rule,
$p_{(j|i)} = p_j p_{(i|j)} / p_i$.
The state of the $i$'th generalized coin can be written as ${\cal Q}_i =\sum_j p_{(j|i)}
\cketbra{j}{j}$.
\item Alice views the result $\cket{j}$ obtained by throwing her $i$'th 
generalized coin,
so that now she knows both the result $\cket{i}$ and the resulting $\cket{j}$ (sampled
by throwing her
generalized coin).  
\item Alice ``forgets'' $i$, to be left with the result $\cket{j}$.
Namely, she ``forgets'' which generalized coin she used.
\end{enumerate}

Let us assume that Alice is left with the result $\cket{j}$. 
What can we now learn from this protocol?
After the first step, Alice has the result $\cket{i}$ with probability 
$p_i$. After the third step, Alice has the result  
$\cket{j}$ with probability $ p_{(j|i)} $. Actually, she knows both $\cket{i}$
and $\cket{j}$. 
The probability of this event is $p_{ij} = p_i p_{(j|i)}$.
After ``forgetting'' the outcome $i$, 
the probability of Alice being left with the result $\cket{j}$ is given by
$\sum_i p_{ij}$.

\begin{itemize}
\item CLAIM 1: the probability that Alice is left with the result $\cket{j}$ is 
$p_j$ (note that this is exactly the probability $p_j$
that she wanted to generate for Bob's state ${\cal P}_j$).

PROOF: As we said, 
the probability of Alice being left with the result $\cket{j}$ is given by
$\sum_i p_{ij}$, which equals $p_j$.
\item CLAIM 2:  The state remaining in Bob's hands is 
${\cal P}_j$.

PROOF:
{}From the Bayes rule we get the probability of Bob's result being $\cket{i}$ given
that Alice's result is $\cket{j}$:  $p_{(i|j)} = p_i p_{(j|i)} /p_j$.
However, this means that Bob's state is ${\cal P}_j$.
\item COROLLARY:
Alice succeeded to generate the desired ${\cal P}$-ensemble.
\end{itemize}
For a fully-detailed example: see Appendix~\ref{App:example}.

Three important remarks are now in order.
First, as in the quantum case, also here --- given the prepared state, and 
Alice's choice of the 
${\cal P}$-ensemble, all four steps of the protocol can be combined 
into a 
single operation which we call a ``generalized measurement'', 
analogous to the 
generalized quantum measurement (POVM).
Second, the above analogy is based on the following special case of
the quantum POVM: The case in which the POVM elements are of rank larger than 
one (this is the quantum analogue of ``forgetting'' the result $i$).
Third, if we only use rank-1 POVM, the quantum case
(which can make use of nonorthogonal states) is still very interesting,
as the number of outcomes, namely $|J|$, 
can be larger than the size of the original reduced state which is $|I|$. 
In contrast, if we are only allowed to use states diagonal in the computation 
basis then using rank-1 POVM leads to the trivial case of just measuring
$\cket{i}$. In this respect, it is clear that the classical nonlocality is
trivial unless the operation of ``forgetting'' is used (namely rank larger than
one for the POVM).

\section{Discussion}

In this work we analyzed in detail a classical teleportation scheme suggested by 
Cohen~\cite{Cohen03}, and by Meyer and Gour~\cite{GM05}, and we proposed
a classical nonlocality argument which resembles the quantum nonlocality
arguments of EPR and HJW.

We think that this work sheds a new light on the question of what
quantum nonlocality really means.  
While quantum nonlocality without entanglement was already
discovered~\cite{qne,upb1} and became the topic of extensive research, 
the topic of nonlocality without any quantumness is offered here
(following~\cite{Cohen03,GM05}), probably for the first time.

We leave open the question of whether such a classical nonlocality protocol
can lead to interesting and useful classical information processing protocols.

\section{Acknowledgement}

I thank Gilad Gour and Eli Biham for interesting and enlightening discussions,
and I thank Amit Hagar for organizing a conference entitled
``What is quantum in quantum computing'' which inspired me to think
about the related question 
``What is quantum in quantum nonlocality''.  
I also thank Ilana Frank-Mor, Doron Tsur, and Dan Kenigsberg for 
carefully editing this manuscript.

This research was supported in part by the Israeli MOD Research and Technology
Unit, and by Israel Science Foundation --- F.I.R.S.T (grant \#4088104).

\paragraph{A Special Acknowledgement}

I thank the late Asher Peres for his wonderful 
contributions to quantum information
processing and for significantly inspiring both my research 
and my understanding of what constitutes good research.

Asher was always fascinated by the differences between quantum physics 
and classical physics, specifically by entanglement and nonlocality.
I think that the only open question he had regarding quantum teleportation
(that he co-authored, as is well known) was the following:
``Does the fact that the teleported state is {\em unknown}  tell us 
something new about quantum physics?''  
I believe that providing definitions of classical teleportation
and classical nonlocality and looking at their connections
to their quantum counterparts is a research direction which helps to 
demystify quantum teleportation and nonlocality,
and hence is particularly appropriate for this special issue dedicated to Asher's memory.


\appendix

\section{A fully-detailed example}
\label{App:example}

This example is solved here to clarify the protocol:
Let Alice and Bob's state be 
$${\cal P}_{\rm fully\mbox{-}correl}= 
\frac{11}{32} \cketbra{0}{0}_{\rm Alice} 
\otimes \cketbra{0}{0}_{\rm Bob} 
+ \frac{21}{32} \cketbra{1}{1}_{\rm Alice} \otimes 
\cketbra{1}{1}_{\rm Bob} $$
so that Bob's reduced state is 
${\cal P}_{\rm Bob} = 
\frac{11}{32} \cketbra{0}{0} + \frac{21}{32} \cketbra{1}{1} $.
We write Bob's reduced state 
using the basis states $\cket{i}$ as
a vector of norm 1,
$$\mathcal{P}_{\rm Bob} = \left\{  \frac{11}{32}  ;
 \frac{21}{32}
\right\} \ .
$$

Let the desired probability distributions that Alice would like to 
generate at Bob's hands be 
$\mathcal{P}_{j=0} = \left\{  \frac{1}{2}  ;
 \frac{1}{2}
\right\}
$ with probability $p_{j=0} = \frac{1}{2}$,
$\mathcal{P}_{j=1} = \left\{  \frac{1}{4}  ;
 \frac{3}{4}
\right\}
$ with probability $p_{j=1} = \frac{1}{4}$,
and 
$\mathcal{P}_{j=2} = \left\{  \frac{1}{8}  ;
 \frac{7}{8}
\right\}
$ with probability $p_{j=2} = \frac{1}{4}$,
The probabilities are chosen such that 
$\mathcal{P}_{Bob} = \sum\limits _{i=0,1} p_i |i)(i|_B$
with $p_i = \sum\limits _{j=0,1,2} p_j p_{(i|j)}$. Then,  
$p_0 = \frac{1}{2}\frac{1}{2} + \frac{1}{4}\frac{1}{4} +
\frac{1}{4}\frac{1}{8} = \frac{11}{32}$,
and
$p_1 = \frac{1}{2}\frac{1}{2} + \frac{1}{4}\frac{3}{4} +
\frac{1}{4}\frac{7}{8} = \frac{21}{32}$.

Now, depending on the state 
$\cket{i}$ Alice observes at step 1 of the protocol, 
she chooses one of the two generalized 
coins 
$\mathcal{Q}_{i=0} = \left\{  \frac{8}{11} ;
 \frac{2}{11} ;   \frac{1}{11}
\right\}
 $ or   
$\mathcal{Q}_{i=1} = \left\{  \frac{8}{21}  ; 
 \frac{6}{21} ;   \frac{7}{21}
\right\}
 $,
with $p(j|i)$ calculated using Bayes rule.
Then, after throwing the appropriate generalized coin,
$\mathcal{Q}_{i}$, 
Alice obtains a result $\cket{j}$.

By ``forgeting'' $i$, she can now predict Bob's state to be
$\mathcal{P}_j$, and that result is obtained with probability
$p_j$.

\end{document}